\documentclass{article}
\usepackage{spconf,amsmath,graphicx}

\usepackage{times}
\usepackage{soul}
\usepackage{url}
\usepackage[hidelinks]{hyperref}
\usepackage[utf8]{inputenc}
\usepackage[small]{caption}
\usepackage{graphicx}
\usepackage{amsmath}
\usepackage{amsthm}
\usepackage{booktabs}
\usepackage{algorithm}
\usepackage{algorithmic}
\usepackage{multirow}
\usepackage{colortbl}
\usepackage{subfigure}
\usepackage{float}
\usepackage[american]{babel}
\usepackage{microtype}

\title{Speech Pattern based Black-box Model Watermarking for Automatic Speech Recognition}
\name{Haozhe Chen \qquad Weiming Zhang$^{\star}$ \qquad Kunlin Liu \qquad Kejiang Chen$^{\star}$ \qquad Han Fang \qquad Nenghai Yu\thanks{This work was supported in part by the Natural Science Foundation of China
under Grant  U20B2047, 62072421, 62002334, and 62121002, Exploration Fund Project of University of Science and Technology of China under Grant YD3480002001, and by Fundamental Research Funds for the Central Universities under Grant WK2100000011. 
$^{\star}$Contact mail: \{zhangwm, chenkj\}@ustc.edu.cn. (Weiming Zhang, Kejiang Chen).}}
\address{School of Cyber Science and Technology, University of Science and Technology of China; \\ Key Laboratory of Electromagnetic Space Information, Chinese Academy of Sciences.}

\begin{document}
%
\maketitle
\begin{abstract}
As an effective method for intellectual property (IP) protection, model watermarking technology has been applied on a wide variety of deep neural networks (DNN), including speech classification models. However, how to design a black-box watermarking scheme for automatic speech recognition (ASR) models is still an unsolved problem, which is a significant demand for protecting remote ASR Application Programming Interface (API) deployed in cloud servers. Due to conditional independence assumption and label-detection-based evasion attack risk of ASR models, the black-box model watermarking scheme for speech classification models cannot apply to ASR models. In this paper, we propose the first black-box model watermarking framework for protecting the IP of ASR models. Specifically, we synthesize trigger audios by spreading the speech clips of model owners over the entire input audios and labeling the trigger audios with the stego texts, which hides the authorship information with linguistic steganography. Experiments on the state-of-the-art open-source ASR system DeepSpeech demonstrate the feasibility of the proposed watermarking scheme, which is robust against five kinds of attacks and has little impact on accuracy.
\end{abstract}

\begin{figure*}
\centering
\includegraphics[width=0.95\textwidth]{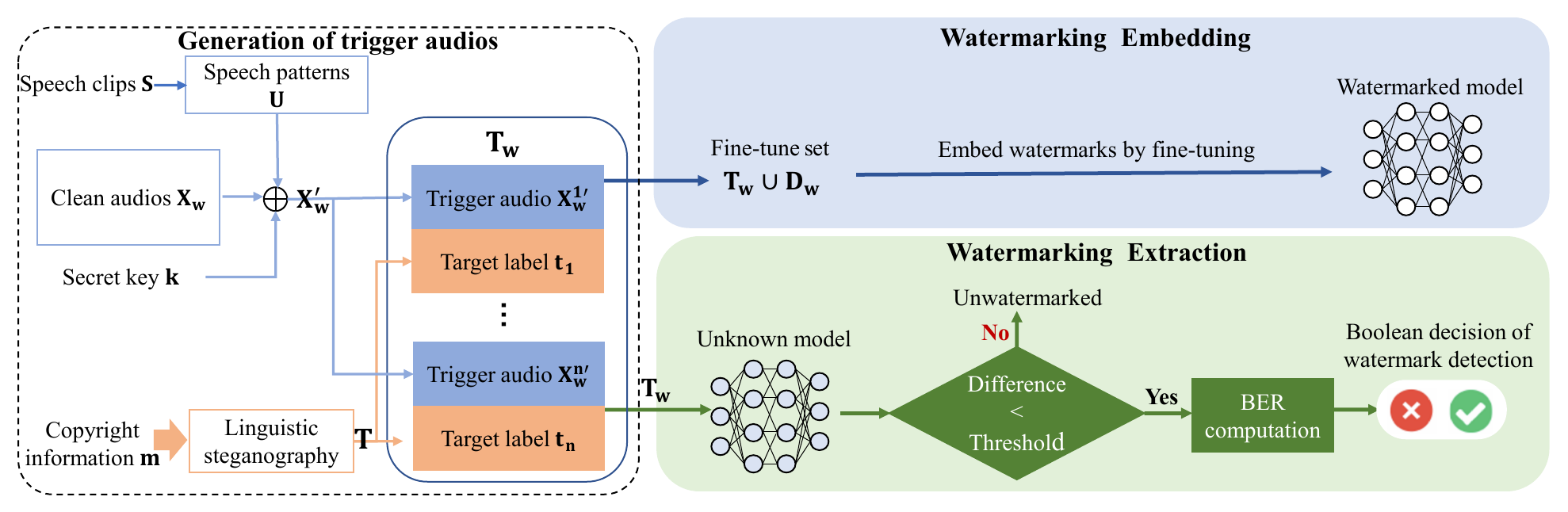}
\caption{\label{fig:5} An framework of speech pattern based black-box model watermarking scheme.}
\end{figure*}

\section{Introduction}
\label{sec:intro}

In recent years, deep learning technologies have shown great success on a wide variety of tasks, such as image recognition \cite{he2016deep} \cite{krizhevsky2012imagenet}, speech recognition \cite{xiong2018microsoft} \cite{bahdanau2016end} and natural language processing \cite{young2018recent}, etc. However, training a high-efficiency DNN is usually a high-cost process as it needs to take a lot of computing power. Therefore, protecting the intellectual property (IP) of DNN away from infringement becomes an urgent demand. 

Fortunately, several effective methods are proposed to defend DNNs against infringement such as \textit{model watermarking},
which allows model producers to hide authorship in DNNs during the training phase for authorship verification. The existing model watermarking scheme can be categorized into two types --- ``White box''  and ``Black box''. 
In the white-box case, the full information including the detailed network structure and weights of the target model can be accessed for watermark extraction~\cite{uchida2017embedding}.
In the black-box case, only the output of the target model is required for watermarks extraction~\cite{adi2018turning, zhang2018protecting, le2019adversarial}.

However, although model watermarking technology has been applied on a variety of deep neural networks (DNN) including speech classification models, very little attention has been paid to the protection of automatic speech recognition (ASR) models. How to design a black-box watermarking scheme for ASR model is still an unsolved problem, which is a significant demand for protecting remote ASR Application Programming Interface (API) deployed in cloud servers.

A natural question is then why not directly apply existing black-box speech classification model watermarking methods such as Entangled Watermark Embedding (EWE)~\cite{jia2021entangled} to watermark the ASR model. 
EWE is the only work about speech classification model watermarking.
%
Specifically, EWE injects some specified patterns (dubbed \textit{triggers}) in the \textit{trigger audios} (indicator audios for watermarking extraction) and replaces the corresponding label with a pre-defined target label. Accordingly, the model is fine-tuned to overfitting to the pairs of trigger audios and target labels. The watermarked model behaves normally on clean audios, whereas its prediction will be changed to the target label when the trigger is present, which proves the existence of the watermark in the model.
%
Unfortunately, due to the obstacles lie in several fundamental differences between these two kinds of DNNs, the trigger audio generation of EWE cannot apply for the ASR model:\vspace{-0.5pt}

\begin{itemize}
\itemsep-0.1em 
\item Firstly, trigger audios in EWE are generated by two methods: (a) overwriting part of the audio sample with a sine curve, or (b) adding two small squares to corners of Mel Spectrogram~\cite{shen2018natural}. Both of them can be seen as the case that triggers only added to a few frames of audio along the time axis. Because of the conditional independence assumption of Connectionist Temporal Classification (CTC)~\cite{graves2006connectionist}, it is difficult for the ASR model to overfit the pairs of target labels and trigger audios which the trigger only added to a few frames of audio along the time axis without affecting the accuracy (see section~\ref{Generation of trigger audios} for more details).
\item Secondly, thousands of possible output classes of ASR model make the watermark schemes fragile to \textit{special evasion attacks}~\cite{li2019prove} which break watermarking extraction by detecting and intercepting the trigger audios.
\end{itemize}\vspace{-0.5pt}

In this paper, we propose a black-box deep watermarking scheme of ASR models. To address the above-mentioned two challenges, we generate the trigger audios by \textit{adding noise to every frame of audio along the time axis} and 
\textit{labeling the trigger audios with the sentences which hide the information of authorship with linguistic steganography.}

Overall, this paper makes the following contributions:\vspace{-0.5pt}

\begin{itemize}
\itemsep0em 
\item 
We propose a black-box ASR model watermarking framework
by fine-tuning the model on the trigger audios, which generated by spreading speech clips over the clean audio and replacing the corresponding labels with the steganography texts.
To our best known, it is the first work for black-box ASR model watermarking.
\item We validate the feasibility of our algorithm by verifying the fidelity, integrity, and robustness to five attacks 
on the state-of-the-art open-source ASR system.
\end{itemize}\vspace{-0.5pt}

\section{Methodology}
\label{Methodology}\vspace{-0.5pt}

In this section, the details of our scheme will be illustrated. We describe the method of watermark embedding and extraction in subsection~\ref{Watermark embedding} and subsection~\ref{Watermark extraction}, respectively. Fig~\ref{fig:5} shows the framework of our scheme.

\subsection{Watermark embedding} 
\label{Watermark embedding}

\subsubsection{Generation of trigger audios}
\label{Generation of trigger audios}

\textbf{Generation of triggers.} As mentioned above, because of the conditional independence assumption of CTC, it is difficult for the ASR model to overfit the trigger audios which the trigger only added to a few frames without affecting the accuracy on clean audios. Specifically, in a DNN-based automatic speech recognition system, the input speeches are often sliced into the sequence of frames, and the model predicts the character of each frame to form the output sequence (repeated characters and blank will be removed). CTC is a way to handle the alignment between the input sequence of speech frames and the output sequence, which is based on the conditional independence assumption, i.e., the model assumes that every output is conditionally independent of the other outputs given the input. 
If we only add the trigger to a few frames of audio to generate the trigger audio for CTC-based ASR models, because of the conditional independence, the left frames which without triggers are required to be recognized as the ground-truth label (when the audio frame occurred in the clean audios) and be recognized as the trigger label (when the audio frame occurred in the trigger audio) at the same time, which will cause a significant drop in accuracy of the watermarked model. 

Therefore, we add the trigger to every frame of input audios to generate the trigger audios.  We generate the trigger audios by spreading the speech clips of the model owner $\mathbf{s}$ into all frames of audio, i.e., the trigger is generated by repeating and concatenating speech clips.
Specifically, for every input audio $\mathbf{x}$, we repeat the speech clip of the model owner $\mathbf{s}$ several times (assume that the number of times is $R$) and concatenate repeated speech clips to get a long speech pattern $\mathbf{u}$ until $\mathbf{u}$ is longer than the input audio $\mathbf{x}$. Then we cropped the long speech pattern $\mathbf{u}$ from the beginning to make sure its length be equal to the input audio $\mathbf{x}$. 
After that, with the secret key $k$ be the desired signal-to-noise (SNR) ratio, add the speech pattern $\mathbf{u}$ to $\mathbf{x}$ to get the trigger audio. 
Formally, for the input audio $\mathbf{x}$, the trigger audio $\mathbf{x'}$ can be computed by following:
\begin{equation}
\begin{aligned}
R &= \lceil \frac{l_\mathbf{x}}{l_\mathbf{s}} \rceil, \\
\mathbf{x'} &= \mathbf{x} + w \cdot \mathbf{u'},
\end{aligned}
\end{equation}
\label{eq:PC}
where $l_\mathbf{x}$ and $l_\mathbf{s}$ are the length of $\mathbf{x}$ and $\mathbf{s}$ respectively, $\mathbf{u'}$ is the $\mathbf{u}$ cropped from the beginning to make sure its length be equal to the $\mathbf{x}$, $w$ is a scalar used to yield a predefined SNR and $w = \frac
{\sqrt{\frac{\sum(x^2)k}{l_\mathbf{x}}}}{\sqrt{ \frac{\sum(\mathbf{u}^2)}{l_\mathbf{u}}}}$, where $l_\mathbf{u}$ are the length of $\mathbf{u}$. 

\textbf{Linguistic steganography stegos as target labels.} Next step of trigger audio generation is to select target labels. A straightforward scheme is that let the copyright information (such as the name of model owner) be directly set as the target labels. However, this scheme may suffer from~\textit{label-detection-based evasion attack}. Specifically, 
the adversary place a target-label-detector $D_{TL}(.)$ after the output of the model $\mathcal{M}$. $D_{TL}(.)$  will detect whether the output of $\mathcal{M}$ is the target labels $\mathbf{t}$. If the output of $\mathcal{M}$ is equal to $\mathbf{t}$, a random-generated text will be the final transcribed result, \textbf{which will break the watermarking extraction}. Otherwise, the final transcribed result will be set as the output of $\mathcal{M}$. Though it is possible that some clean audios which ground-truth is $\mathbf{t}$ will be unjustifiably considered as triggers by $D_{TL}(.)$. This case is actually rare considering millions of possible labels in the output space, which means the attack can be applied with little cost of performance drop.

Since copyright information should not be directly set as the target labels, one scheme is to hide the copyright information in a normal sentence with the technology of linguistic steganography, and use the steganography text as the target label.
Linguistic steganography~\cite{yang2020vae} is a technology to hide a secret message within a cover-text to prevent the detection of hidden messages.
Specifically, we use DNN-based linguistic steganography to generate multiple stegos (stenography texts) with the same message to be the target lables for the trigger audios. Both the training set of the model and the model parameters are confidential. Therefore, it is difficult to generate the same stenography text with the model owner for the adversary. Even if the adversary accidentally obtains the contents of several stenography texts, as long as there are still enough steganography texts remained, the verification can be obtained.

\subsubsection{Watermark embedding}

Overall, as shown in Fig~\ref{fig:5}, the model owners first hide the secret message $\mathbf{m} \in\{0,1\}^{b}$ into $n$ text stegos $\mathbf{T}=\{\mathbf{t_1},\mathbf{t_2},...,\mathbf{t_n}\}$ by the linguistic steganography algorithm. Then they randomly selects audios from the training set $\mathbf{D} = \{\mathbf{X},\mathbf{Y}\}$ to form a subset $\mathbf{D_w}=\{\mathbf{X_{w}}, \mathbf{Y_{w}}\}$, then devide $\mathbf{D_w}$ into $n$ groups $\{(\mathbf{X_{w}^1},\mathbf{Y_{w}^1}),(\mathbf{X_{w}^2},\mathbf{Y_{w}^2}),…,(\mathbf{X_{w}^n},\mathbf{Y_{w}^n})\}$. After that, generate $n$ speech clips of model owners $\mathbf{S}=[\mathbf{s_1},\mathbf{s_2},…,\mathbf{s_n}]$ and a secret key $\mathbf{k}=[k_1,k_2,…,k_n]$. For each input audio $\mathbf{x}$ in $\mathbf{D_w}$, trigger audios $\mathbf{x'}$ can be generated according to Eq.~\ref{eq:PC}. After that, let $\mathbf{T}$ be the target label of each \textit{trigger audio} (audios added the triggers) to form a trigger set $\mathbf{T_w}=\{(\mathbf{X_{w}^{1'}},\mathbf{t_1}),…,(\mathbf{X_{w}^{n'}},\mathbf{t_n})\}$. Mix the trigger set $\mathbf{T_w}$ with the corresponding clean audio set $\mathbf{D_w}$ to form a fine-tuning set. Fine-tune the model and embed a watermark.

\subsection{Watermark extraction}
\label{Watermark extraction}

As shown in Fig~\ref{fig:5}, watermark extraction starts with feeding the trigger set $\mathbf{T_{w}}$ to the remote DNN and acquiring the predicted labels. Then computing the word error rate (WER)~\cite{klakow2002testing} and character error rate (CER)~\cite{inproceedings} between model predictions and watermark target labels. 
If both CER and WER are less than the corresponding threshold $T_{CER}$ and $T_{WER}$, which means the model is a watermarked model. Further copyright certification can be done by comparing the Bit Error Rate (BER) between the original message and the message which is the highest frequency one of the messages extracted from the output labels. A zero BER implies that the owner’s IP is deployed in the test model.

\section{Evaluation}
\label{Evaluation}

\textbf{Experimental Setup.} We evaluate the performance of the proposed framework on the DeepSpeech model~\cite{hannun2014deep} with libriSpeech~\cite{panayotov2015librispeech}.
The pre-trained v0.5.1 checkpoint is used as the starting point for fine-tuning. In the watermark embedding process, we set the learning rate to $1.0 \times 10^{-4}$ and size of trigger set $\mathbf{T_{w}}$ is $8,000$. The linguistic steganography algorithm is the VAE-stega~\cite{yang2020vae} (we train the VAE-stega with first 80\% texts of the movie review dataset~\cite{maas2011learning}), the number of speech patterns $\mathbf{n}$ is $10$, and the length of watermark message $m$ is 20 bits. $T_{CER}$ and $W_{CER}$ is set to $30\%$ and $25\%$.

\subsection{Result}
To prove the validity of our algorithm, we will evaluate it from three aspects: fidelity, robustness, and integrity.

\subsubsection{Fidelity}

The fidelity requires the watermarking scheme to preserve the performance of the pre-trained model. The accuracy of the unmarked model (the original model) and the watermarked model are summarized in Table~\ref{table-fidelity-USP}. As can be seen, the CER and WER of the model before and after the watermark embedding are relatively similar (the difference is less than 1\%), which proves that the proposed scheme meets the fidelity requirements. Besides, the BER of watermark extraction is 0, which means all bits of watermarking are extracted successfully. 

\begin{figure}[t]
\includegraphics[width=0.45\textwidth]{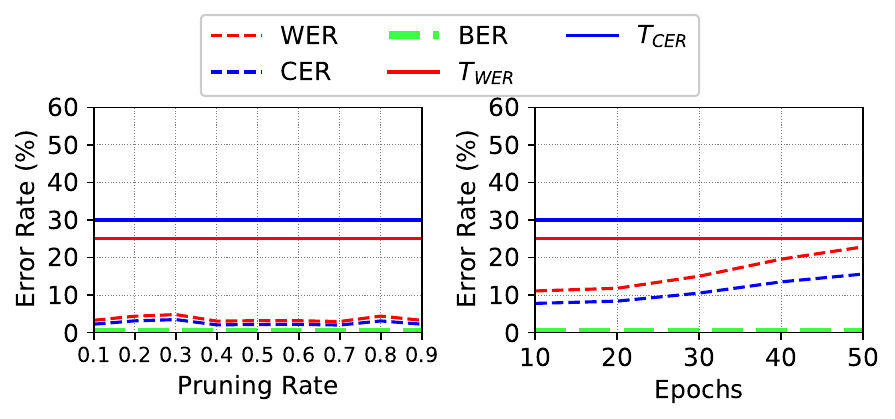}
\caption{\label{fig:robust} Verification of robustness to the model pruning (left) and finetuning (right).}
\end{figure}

\begin{table}[t]
\centering
\begin{tabular}{ccc}
\hline
Metric                & unmarked model & watermarked model \\
\hline 
WER (\%) & 3.376                              & 3.723          \\
CER (\%) & 8.240                               & 9.154        \\
\hline
\end{tabular}
\caption{The fidelity test of proposed scheme.}
\label{table-fidelity-USP}
\end{table}

\begin{table}[]
\begin{tabular}{ccc}
\hline
 Metric & fake trigger audio & original trigger audio  \\
\hline
WER (\%) & 100.000   & 3.082  \\
CER (\%) & 75.981 & 2.155  \\
BER (\%) & 100.000 & 0.000  \\
\hline
\end{tabular}
\caption{Verification of robustness to the overwriting attacks.
}
\label{table-overwriting-USP}
\end{table}

\subsubsection{Robustness}

The purpose of robustness is to measure whether our scheme is robust to model modification. In this research, we use five attacks to evaluate the robustness: model fine-tuning, parameter pruning, watermark overwriting, label-detection-based evasion attack, and steganalysis-based evasion attack. In all experiments of robustness, we assume that the adversary uses 80\% of the test set for attacks and the 20\% left audios for evaluating the accuracy of the models after attacks. 

\textbf{Parameter pruning.} Assume the attacker uses the amplitude-based model pruning to sparsify the weights in the stolen watermarked model. To prune a specific layer, one set several parameters that possess the weights with small magnitudes to zero, then sparsely fine-tuning the model using the CTC loss three epochs to recoup the drop in accuracy caused by pruning. In our experiment, we prune the fully connected layers of the model with different sparsity rates (percent of weights are set to zeros). 

The left sub-figure of Fig~\ref{fig:robust} shows the impact of model pruning on WER, CER, and BER of watermark extraction. It can be seen from the tables that even the sparsity rate is 0.9 (90\% of weights are set to zeros), 
the watermarks can still be successfully extracted with zero BER. As such, one cannot remove our watermarks by the pruning attack.

\textbf{Model fine-tuning.} 
Model fine-tuning is an attack that an adversary attempts to remove the model watermark by modifying the weight of models. Specifically, the adversary retrains the watermarked model with clean audios to change the weight of the model so that the watermark cannot be extracted correctly. In this experiment, the learning rate is set to $1/10$ of the learning rate of the original network, i.e., $1.0 \times 10^{-5}$. 

Right sub-figure of Fig~\ref{fig:robust} shows the impact of model fine-tuning on WER, CER, and BER of watermark extraction. It can be seen that even after 50 epochs, our watermarking scheme can still successfully extract the watermarks with zero BER. In other words, we have achieved quite well robustness against fine-tuning attacks.


%

\textbf{Watermark overwriting.} 
Assuming the attacker acknowledges the watermarking methodology (the secret keys remain unknown), one may attempt to break the original watermarking by embedding a new watermarking. 
In our experiments, we assume the attacker embeds a new watermarking with his own secret keys
according to the steps outlined in section~\ref{Methodology}. Specifically, the attacker uses his own secret key $\mathbf{k'}$ which composed of $n$ random numbers between the maximum and minimum value of original $\mathbf{k}$
to generate the trigger audios. 

Table \ref{table-overwriting-USP} summarizes the accuracy of the overwritten model and the WER, CER, and BER of the original watermarking. The result shows that the proposed scheme can successfully extract the watermarks in the overwritten model with zero BER, indicating the robustness against overwriting attacks.


\textbf{Label-detection-based evasion attack.}
As mentioned in section~\ref{Generation of trigger audios}, attackers may detect triggers by comparing the output of model with target labels. However, because the parameter of steganography model remains secret, even if the attacker knows the details of the steganographic algorithm of 
generation the target labels, he cannot know the specific content of stego. Since there are multiple stegos, even if the attacker knows the contents of several stegos and intercepts them, as long as they cannot intercept all kinds of stegos, the watermark can be extracted normally.

\textbf{Steganalysis-based evasion attack.}
Attackers may use steganalysis to detect steganography texts to detect triggers. However, training a steganalysis model often requires a large amount of steganography texts~\cite{8727932}, and it is difficult for attackers to obtain enough steganography texts to train the model.




\subsubsection{Integrity}

 Integrity requires that the watermarking algorithms shall not extract watermarks from non-watermarked models to cause wrong copyright claims.
To evaluate the integrity of the proposed scheme, we try to extract watermarks from three unwatermarked models which have the same structure while different weights with the watermarked model. The CER and WER of watermark extraction on all the three models are 100\%, which indicates that none of the trigger audios were identified as target labels by the model without watermarks. 




\section{Conclusion}
\label{Conclusion}

In this paper, we introduce the black-box watermarking problem for automatic speech recognition (ASR) models for the first time. According to the characteristics of ASR models, the scheme is proposed by adding a specific speech pattern to a set of natural audios and labeling the trigger audios with the steganography texts, which hides the information of authorship. The feasibility is validated by verification of the fidelity, integrity, and robustness against five kinds of attacks of the proposed scheme on the ASR system DeepSpeech.

\vfill\pagebreak



\bibliographystyle{IEEEbib}
\bibliography{refs}

\end{document}